\documentclass[aps,prl,preprint,superscriptaddress]{revtex4-2}
\usepackage{multirow}
\usepackage{amsmath}
\usepackage{amssymb}
\usepackage{graphicx}
\usepackage{dcolumn}
\usepackage{bm}
\UseRawInputEncoding

\usepackage{soul,color,xcolor}

\usepackage[colorlinks,
colorlinks=true,
linkcolor=blue,
filecolor=blue,
urlcolor=blue,
citecolor=cyan,anchorcolor=blue,citecolor=blue]{hyperref}

\begin{document}


\title{Joule heating and electronic Gurzhi effect in hydrodynamic differential transport in an electron liquid}

\author{Yi Wang}
\affiliation{State Key Laboratory of Semiconductor Physics and Chip Technologies, Institute of Semiconductors, Chinese Academy of Science, Beijing, 100083, China}
\affiliation{College of Material Science and Opto-Electronic Technology, University of Chinese Academy of Sciences, Beijing 100190, China}

\author{Shu-Yu Zheng}
\affiliation{Beijing National Laboratory for Condensed Matter Physics, Institute of Physics, Chinese Academy of Sciences, Beijing 100190, China}

\author{Li Lu}
\affiliation{Beijing National Laboratory for Condensed Matter Physics, Institute of Physics, Chinese Academy of Sciences, Beijing 100190, China}

\author{Kai Chang}
\affiliation{Center for Quantum Matters, School of Physics, Zhejiang University, Hangzhou 310058, China}

\author{Chi Zhang}
\altaffiliation{Electronic address: zhangchi@semi.ac.cn}
\affiliation{State Key Laboratory of Semiconductor Physics and Chip Technologies, Institute of Semiconductors, Chinese Academy of Science, Beijing, 100083, China}
\affiliation{College of Material Science and Opto-Electronic Technology, University of Chinese Academy of Sciences, Beijing 100190, China}
\affiliation{CAS Center for Excellence in Topological Quantum Computation, University of Chinese Academy of Sciences, Beijing 100190, China}


\begin{abstract}
We perform a differential resistance study in the hydrodynamic regime of electron liquid in GaAs/AlGaAs quantum wells.
At zero magnetic field ($B$) a Lorentzian profile occurs in the nonlinear transport driven by a U-turn (ac) current loop, in (ac + dc) measurements a minimum deepens with the external dc current bias ($j_{dc}$).
Our analysis shows that the observed electronic transport valley induced by $j_{dc}$ is attributed to Joule heating effect on the electron temperature ($T_{e}$) of electron liquid.
Quantitatively, we demonstrate that the viscosity resistivity ($\Delta \rho$) is proportional to $T^{-2}$ and is consistent with the dc-current induced electronic Gurzhi effect in various configurations of measurement.
\end{abstract}

\maketitle


\section{Introduction}\label{sec:introduction}

High-mobility two-dimensional electron gas (2DEG) has been a reliable platform for investigating various novel phenomena in transport study \cite{PhysRevLett.45.494, PhysRevLett.48.1559}.
In the past few years, problems concerning electrons or quasiparticles behaving like viscous fluids have raised a lot of interest \cite{sulpizio_visualizing_2019, li_nonlocal_2022, raichev_superballistic_2022, engdahl_driving_2024, estrada-alvarez_superballistic_2025, patricio_magnetic_2024}.
The viscosity of electron liquid is an approximate approach to deal with complicated electron-electron (e-e) interaction in a collective mode \cite{policastro_shear_2001, son_vanishing_2007, andreev_hydrodynamic_2011, khoo_quantum_2020}.
In the study of electron liquids, the relaxation times and the mean free paths (m.f.p.) play key roles in both classical and quantum physics approaches.
In a high-mobility ($\mu \sim (3 - 5) \times 10^6\ \mathrm{cm^2/ V\cdot s}$) 2DEG, the mean free path ($l$) governed by the scatterings from phonon or disorder is much larger than the Hallbar width $w$, meanwhile, the characteristic length of e-e collisions $l_{ee}$ is smaller than $w$ \cite{andreev_hydrodynamic_2011, torre_nonlocal_2015}.
In the regime, transport features are largely dominated by e-e collisions, impurities and sample edges \cite{weiss_electron_1991, bockhorn_magnetoresistance_2014}.
Since the current flow in electron liquid resembles the Poiseuille flow in conventional fluid, where the shear viscosity $\eta \sim v_{F} l_{ee}$, anomalies occur in the nonlinear or nonlocal transport of electron fluids \cite{hoyos_hall_2012, principi_bulk_2016, holder_unified_2019, gorlan_boundary_2019}.
The theory has successfully described characteristics in viscous electron fluid with the hydrodynamic equations of charge-compensated fluids at low $B$ \cite{andreev_hydrodynamic_2011, alekseev_negative_2016, pellegrino_nonlocal_2017}.

However, in hydrodynamic transport studies \cite{krishna_kumar_superballistic_2017, gusev_viscous_2018, levin_vorticity-induced_2018, berdyugin_measuring_2019, sulpizio_visualizing_2019}, the effects from biased $E$-field on the electron liquids are still open questions \cite{PhysRevB.107.195406}.
By introducing an external dc bias,  the electrons transform into another equilibrium.
Due to the nonlinear inertial item of Navier-Stokes equation, the electron fluid exhibits velocity-dependent transport features.
It is still ambiguous regarding how e-e interaction and edge geometry manipulate the transport process of electron viscous fluid \cite{levin_vorticity-induced_2018}.
Here, we focus on researching the nonlinear (or differential) magneto-resistance under dc (current) bias and sample edge effects in order to investigate novel equilibrium states \cite{levitov_electron_2016, pellegrino_nonlocal_2017, krishna_kumar_superballistic_2017}.

In this study, we examine differential magneto-resistance of viscous electron under high current measurement in high quality 2DEG in a GaAs/AlGaAs quantum well (QW).
In particular, we focus on the nonlinear resistivity of viscous electrons and its effect from the boundary condition of samples.
At around $B = 0$, we observe an unusual dip caused by the increasing second momentum relaxation rate $1/\tau_{2, ee}$, which is in quadratic terms of current density ($j_{dc}$) and can be attributed to the difference between heated electron $T_e$ and lattice $T_l$.

\section{\label{sec:1}METHOD}
We perform our experiments on three high-quality Silicon modulation-doped GaAs/AlGaAs QW wafers grown by molecular-beam epitaxy (MBE).
The high-mobility ($\mu \sim (3-5) \times 10^6\ \mathrm{cm^2/ V\cdot s}$) electron channels in the wafers are about 100 nm beneath the sample surface, and have carrier densities of $n_e \sim (2-3) \times 10^{11}$ cm$^{-2}$ after LED illumination at 1.5 K.
The source-drain channel of all our Hallbar devices is chosen to be $w = 4\ \mu$m in order to confirm $l > w$.
All the Hallbars consist of three consecutive segments of different lengths (5, 10, 5 $\mu$m/8, 15, 8 $\mu$m, see Table I), and eight voltage contacts at both sides of the source-drain channel (as shown in Table \ref{Table1}).
The $w= 4 \mathrm{\mu m}$ wide source-drain channel is chosen to realize hydrodynamic regime electron fluid, and is close to e-e correlation $l_{ee}$ so as to observe phase transition between Ohmic transport and the hydrodynamic transport regime \cite{PhysRevB.104.195415}.

 \begin{table*}
 	\begin{tabular}{|c|c|c|c|c|c|c|}
 		\hline
 		Sample & $n$ (10$^{11}$ cm$^{-2}$)& $\mu$ (10$^6$ cm$^2$/V s)&	Structure&	$w$ ($\mu$m)& $d$ ($\mu$m)&	$L$ ($\mu$m) \\ \hline
 		\multirow{2}{*}{A} &2&2.2	&I	&4	&5	&15\\ \cline{2-7}
 		&2&	2.2&II&	4&	5&	8\\ \hline
 		B&	2.9&4.2	&III&	4&	5&	10\\ \hline
 		C&	1.8&	3&	IV&	3&	0.6&	5.4 \\ \hline
  	\end{tabular}
  	\caption{Parameters and structures of the Hallbar devices. The electron densities $n$ and mobilities $\mu$ of three wafers are listed. And the Hallbar parameters, Hallbar widths $w$, $d$ and length $L$ are shown in the inset of Fig. \ref{Fig1}(a).}
  	\label{Table1}
 \end{table*}

All the measurements are carried out in a cryogen-free VTI cryostat with a base-$T$ of 1.5 K and a vector magnet (8 T / 2 T).
We examine longitudinal $\rho_{xx}$ resistivity by using the 4-terminal probe method via lock-in amplitudes (at $f = 19$ Hz).
We impose a low ac current at 0.1 $\mathrm{\mu A}$ to minimize heating from measurements, and apply an external dc current range from 0.1 $\mathrm{\mu A}$ to 15 $\mathrm{\mu A}$ to probe its effect on the electron liquid.
The external $B$ applied perpendicular to the sample surface (at a range of from -8 to 8 T) is generated by superconducting coils.

Due to the boundary condition-dependent transport of electron fluid, we mainly conduct the transport measurements with two different configurations in our experiments:
U-shape measurement (as shown in Fig. 1(a)) has a driven U-turn current and voltage probes located on the opposite side \cite{torre_nonlocal_2015, ji_negative_2021}.
Ordinary measurement (as shown in Fig. 4(a)) is the standard 4-terminal measurement in which the current flows through the main path (along the $x$-axis of the sample) \cite{shepard_direct_1992, alekseev_viscosity_2020}, and voltage drop is measured between adjacent electrodes.

\section{\label{sec:2}EXPERIMENT RESULTS}
Our experimental results show differential resistance (under dc bias) in hydrodynamic transport, in which $l$ is much larger than $w$, and $l_{ee} < w$.
In order to explore the nonlinear changes in resistivity, a differential resistivity $\rho_{xx}^\ast$ is measured with external currents (ac + dc), in which the superscript ``$\ast$" expresses the differential resistance or differential resistivity.

\begin{figure}[ht]
\includegraphics[width=\linewidth]{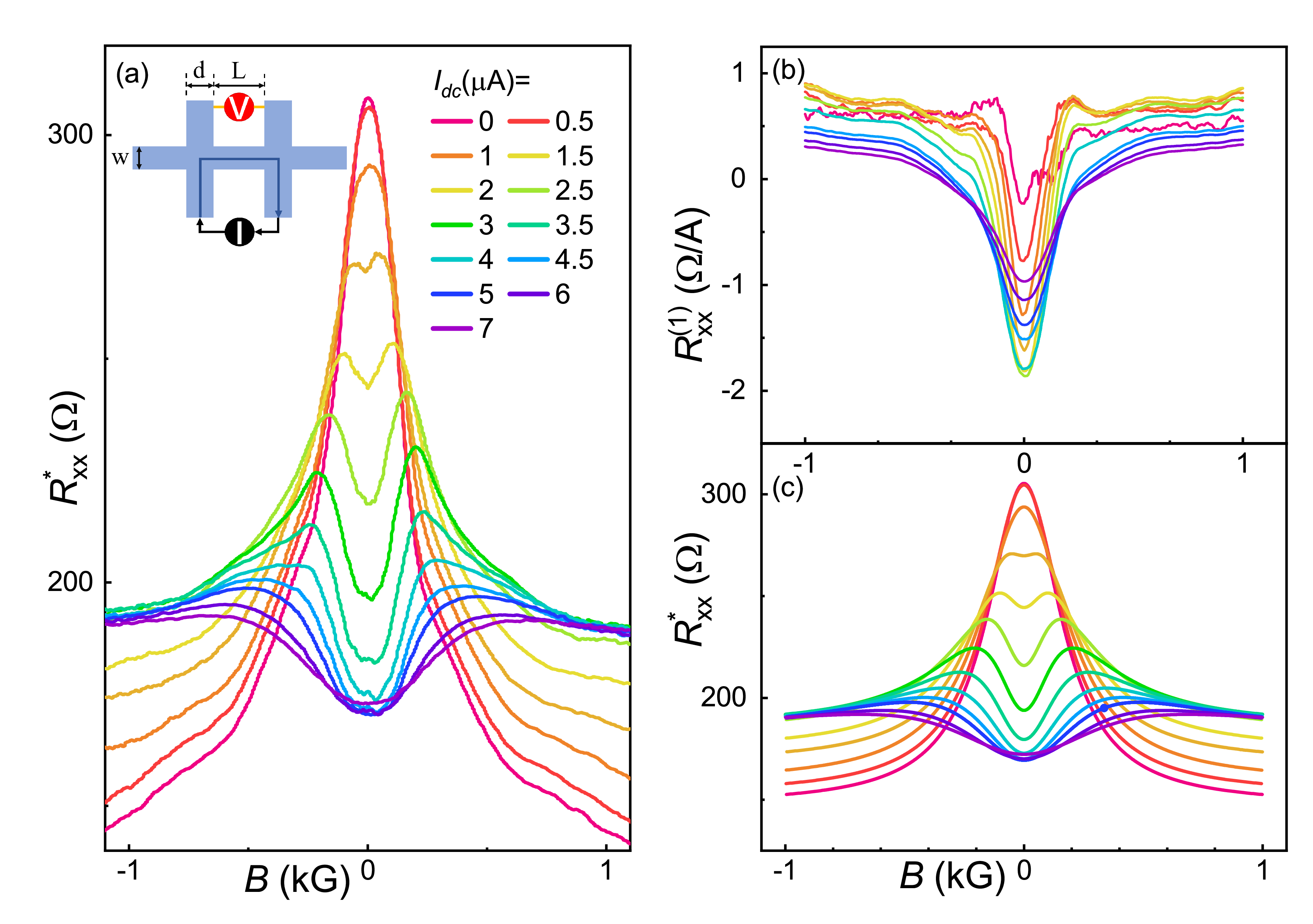}
\caption{\label{Fig1} (Color online).
Panel (a): Differential (longitudinal) resistance $R^{\ast}_{xx}$ measured at 1.6 K using U-shape configuration.
(b): The first derivative of resistivity $R^{(1)}=(R^{\ast}_{xx}(I)-R^{\ast}_{xx}(0))/2I_{dc}$ with respect to dc current.
(c): Fitting curves of the traces in panel (a) via the superposition of two Lorentzian profiles.}
\end{figure}

In Fig. 1 we display differential resistance $R^\ast_{xx}$ as a function of dc current (or current densities).
The external dc currents are applied to Structure I (in the range of 0.5 -7 $\mathrm{\mu A}$) at $\sim 1.5$ K.
All the measurements are carried out under the configuration driven by (ac + dc) U-turn currents, as shown in Panel (a).
At low $B$ (-0.1 $\mathrm{T}$ $<B<$ 0.1 $\mathrm{T}$), a giant negative resistance occurs with a Lorentzian profile \cite{shi_colossal_2014, delacretaz_transport_2017, scaffidi_hydrodynamic_2017}.
When the dc current increases, both structures display two features: (i) The Lorentzian peak broadens and flattens, and eventually disappears.
(ii) An additional dip appears around $B = 0$, and grows stronger, but vanishes with increasing $I_{dc}$ or $j_{dc}$.
As displayed in Fig. 1(a), a dip at $B = 0$ appears at a dc current of $I_{dc} = 0.5\ \mathrm{\mu A}$, but weakens and vanishes at $I_{dc} > 6 \ \mathrm{\mu A}$.

In hydrodynamic transport, a robust dip with a Lorentzian profile occurs in magneto-resistance, which is attributed to the semi-classical approach of viscous fluid.
The introduced second momentum relaxation rate ($1/\tau_{2, ee}$) for describing e-e interaction is proportional to the viscosity in electron fluid, which is characterized by the kinematic equation \cite{alekseev_negative_2016}:
\begin{eqnarray}
\frac{\partial\mathbf{v}}{\partial t}=\eta_{\mathrm{xx}}\Delta \mathbf{v}+[(\eta_{\mathrm{xy}}\Delta\mathbf{v}+\omega_{c}\mathbf{v})\times\mathbf{e}_{z}]+\frac{e}{m}\mathbf{E}-\frac{\mathbf{v}}{\tau}
\end{eqnarray}
In Eq.(1), $\mathbf{v}$ refers to the space distribution of the drift velocity of liquid, and $\tau$ is the relaxation time of electrons.
Thereby, the viscosity and resistance tensors can be expressed as follows \cite{alekseev_negative_2016}:
\begin{equation}
    \left\{
\begin{aligned}
   \eta_{xx}=\eta\ \frac{1}{1 + (2 \omega_{c} \tau_{2})^{2}}\\
   \eta_{xy}=\eta\ \frac{2 \omega_{c} \tau_{2}}{1 + (2\omega_{c} \tau_{2})^{2}}
\label{eq0}
\end{aligned}
\right.
\end{equation}
\begin{equation}
    \left\{
\begin{aligned}
   \rho_{xx}&=\rho_{0}^{bulk}\ (1 + \frac{\tau}{\tau^{\ast}}\frac{1}{1 + (2 \omega_{c} \tau_{2})^{2}})\\
   \rho_{xy}&=\rho_{xy}^{bulk}\ (1 - r_{H} \frac{2 \tau_{2}}{\tau^{\ast}} \frac{1}{1 + (2 \omega_{c} \tau_{2})^{2}})
\label{eq1}
\end{aligned}
\right.
\end{equation}
where $\rho_{0}^{bulk} = \frac{m^{*}}{ne^{2}\tau}$ refers to the bulk resistivity, $\tau^{*} = w \left(w+6l_{s}\right)/12\eta$ is the effective relaxation time, $\eta = \frac{1}{4}v_F^2\tau_{2, ee}$ is the viscosity of electron fluid, $r_{H}$ is a numerical coefficient (in the order of magnitude of 1) and $l_{s}$ is the slip length.
Pertaining theoretical descriptions are summarized in Section I of Supplementary Information \cite{alekseev_negative_2016, alekseev_viscosity_2020, qian_lifetime_2005}.
The external $B$ plays an equivalent role in the off-diagonal element, e.g. via the cyclotron motion term of electron $\omega_{c} = eB/m^{*}$.
Since the velocity ($v$) is proportional to its driven $E$-field, linear kinematic equation (1) is effective to characterise steady electron flows with low velocities.
But practically it will be invalid for liquids at large velocities or with a strong nonlinearity.
Based on Eq. (3), we estimate the fundamental parameters at zero dc-current (as shown in Fig. 1(a)): $v_{F} = 1.94 \times 10^5$ m/s, $\tau = 83.8$ ps, $l = 16.2 \ \mu$m, $\tau_{2,ee} = 13.4$ ps and $l_{ee}=2.6\ \mu$m, which satisfies the hydrodynamic relation $l_{ee} <W \ll l$.

According to Eq.s (2) and (3), the viscosity-induced maximum close to $B = 0$ can be characterized by $\Delta \rho = m^{\ast}/n e^{2} \tau^{\ast}$ which decreases as a function of $B$.
Despite the abnormal dip, the envelopes of the differential resistance maintain the Lorentzian profiles.
In the Taylor expansion of the nonlinear resistance expression, the first-order derivative of resistance $R^{(1)}=(R_{xx}(I_{dc})^{*}-R_{xx}(0)^{\ast})/2I_{dc}$ (as shown in Fig. 1(b)) also resembles Lorentzian peaks.
Thus we fit the magneto-resistance curves with the superposition of two Lorentzian envelopes (as shown in Fig. 1(c)).
As illustrated in Fig. 1(b), $R^{(1)}$ displays a transverse from positive to negative values at low $B$ with increasing dc currents.

\begin{figure*}[!ht]
\includegraphics[width=\linewidth]{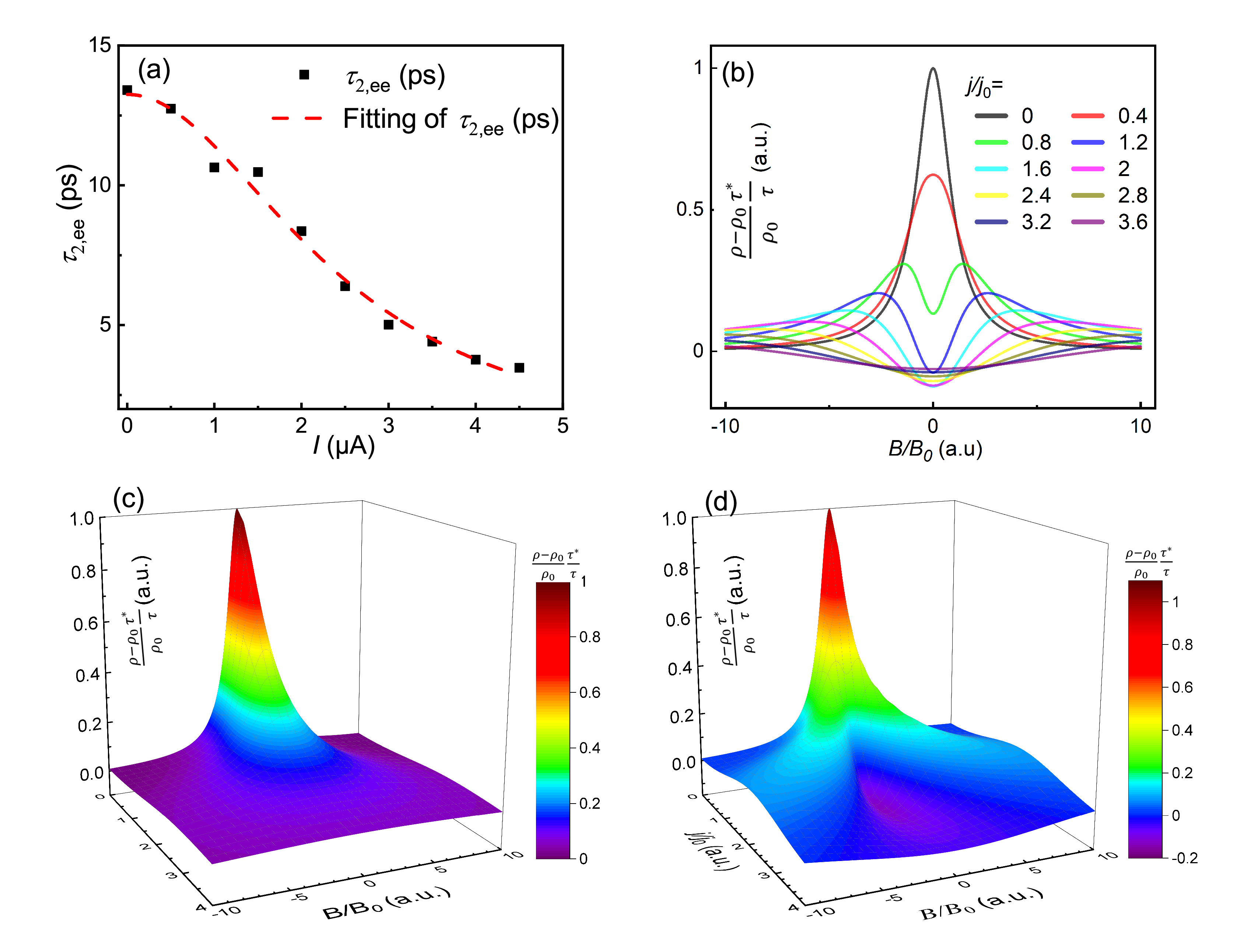}
\caption{\label{Fig2} (Color online).
Panel (a): Lorentzian Fitting parameter: $\tau_{2, ee}$ versus applied dc current.
The red dashed fitting curve shows that $\tau_{2, ee}$ follows the form of Eq.(4), in which $\theta$ is a coefficient.
(b): Normalized differential resistance($d\rho_{xx}/di$) at different dc current.
Both $B$-field and dc current are normalized by parameter $B_0=m\/e\tau_2$ and $i_0$.
(c), (d): 3D calculated contour plot of resistance $\rho$ and differential magnetoresistance $d\rho_{xx}/di$ are obtained via Eq. (\ref{eq6}), respectively.}
\end{figure*}

Our further numerical analysis provides a distinct explanation.
The envelope of Lorentzian peaks conforms to the form $\rho = y_{0} + A/(1 + CB^{2})$, in which the simplified coefficients are: $A = m^{\ast}/n e^{2}\tau^{\ast}$, $C=(2e \tau_{2, ee}/m^{\ast})^{2}$ and $y_{0} = m^{\ast}/n e^{2} \tau$.
Thus we conclude important parameters: $\eta=\frac{1}{4} v_{F}^{2} \tau_{2, ee}$ and $l_{s}=(12 \eta \tau^{\ast}/\omega - \omega)/6$, which are shown in Fig. 3(b) (vs. $j_{dc}$) and Fig. 4(c) (vs. $T$), respectively.
In Fig. 2(a), the relaxation time of $\tau_{2, ee}$ is expressed as follows:
\begin{equation}
   \tau_{2, ee}=\frac{\tau_{2}}{1+ \theta I^2}
\end{equation}
where $\theta$ is merely a defined parameter, and the relaxation rate $1/\tau_{2, ee}$ is approximately parabolic with respect to the dc-current:
\begin{eqnarray}
\frac{1}{\tau_{2, ee} (j_{dc})}=\frac{1}{\tau_{2}}+A_{ee}^{j} j_{dc}^2=\frac{1}{\tau_{2}}+(\frac{j_{dc}}{j_{0}})^2
\label{eq4}
\end{eqnarray}
where $\tau_{2}$ is the $T$-dependent second momentum relaxation time at zero dc-current and $A_{ee}^{j}$ is a coefficient for describing the process with dc currents.
Substituted into Eq. (\ref{eq1}), we obtain the resistivity arising from viscous electrons:
 \begin{eqnarray}
\Delta\rho\left(j\right)&=&\rho\left(j\right)-\rho_0 \nonumber \\
&=&\frac{m}{ne^2\tau^\ast}\ \frac{1}{1+\tau_{2, 0}A_{ee}^j j_{dc}^2+\frac{\left(2\omega_c\tau_{2, 0}\right)^2}{1 + \tau_{2, 0} A_{ee}^{j} j_{dc}^2}}
\label{eq5}
\end{eqnarray}

Thus, we gain a full vision about the function of the dc current in manipulating transport features.
By introducing two normalization parameters $j_{0} = 1/\sqrt{\tau_{2, 0}A_{ee}^j}$, and $B_{0} = m^{\ast}/2e \tau_{2, 0}$, the viscous term of the transport resistance can be simplified as:
\begin{eqnarray}
    \frac{\rho(j_{dc})-\rho_0}{\rho_0}\frac{\tau^{\ast}}{\tau} = \frac{1}{1+(j_{dc}/j_0)^2+(B/B_0)^2/[1+(j_{dc}/j_0)^2]}
    \label{eq6}
\end{eqnarray}
The dc bias-dependent resistivity related to viscosity that is derived from our transport measurements is shown in Fig. 2(b).
Panels (c) and (d) display the 3D contour plot of dimensionless viscous resistivity against the current density $j$ and $B$: Fig. 2(c) shows the resistivity $\rho$ caused by viscous terms (Eq. (\ref{eq6})), and Fig. 2(d) displays the differential resistivity $d\rho/di$.
The viscous resistivity declines with $B$ and dc current.
The robust dip from viscosity has the same order of magnitude as the Lorentzian peak.

\begin{figure}
\includegraphics[width=\linewidth]{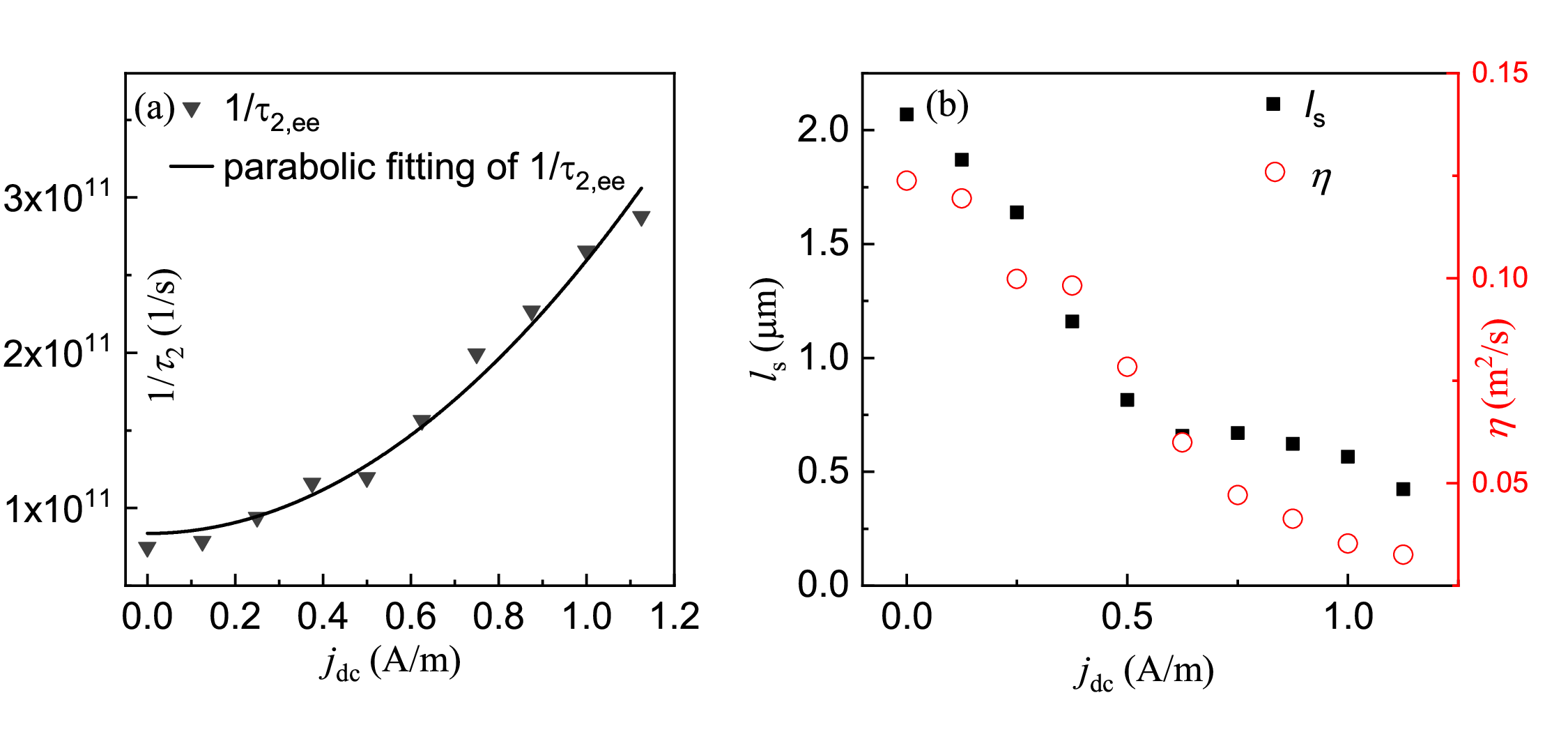}
\caption{\label{Fig3} (Color online).
Panel (a): Second momentum relaxation rate ($1/\tau_{2, ee}$) (black solid triangles) and its parabolic fitting curve versus dc current ($j_{dc}$) in the U-shape measurement of Structure I at 1.6 K.
(b): Slip length $l_s$ (black solid squares) and viscosity $\eta$ (red open circles) obtained from U-shape measurements of Structure-I at 1.6 K.}
\end{figure}

Based on the dc-current-dependent transport, important parameters related to viscous electron are shown in Fig. 3, including viscosity $\eta$, relaxation time $\tau_{2, ee}$ and slip length $l_{s}$.
And our analysis is derived from Eq. (4) with the parameters (at zero $j_{dc}$): $1/\tau_{2, 0} = 8.36 \times 10^{10}$ s$^{-1}$ and $A_{ee}^{j} = 1.7567 \times 10^{11}$ m$^{2}$ s$^{-1}$ A$^{-2}$.
More detailed data and calculated process are displayed in the Supplementary Information (\cite{Supplementary}).
The relaxation rate $1/\tau_{2, ee}$ (red solid triangles) versus $j_{dc}$ follows the parabolic trace which is highlighted by the red solid curve in Fig. 3(a).
Both the slip length $l_s$ and viscosity $\eta$ (in Fig. 3(b)) decrease with the dc current density which is corresponding to the large drift velocity.
As a very strong dc current ($>5\ \mu A$) passes through the structure, the viscous (term) resistivity becomes very small and the electron flow resembles the Ohmic flow.
Since the hydrodynamic regime of electron transport is an transitional process between Ohmic and ballistic transport, external dc-current becomes effective to manipulate the electron fluid to evolve from hydrodynamic regime to Ohmic regime.

\begin{figure*}
\includegraphics[width=\linewidth]{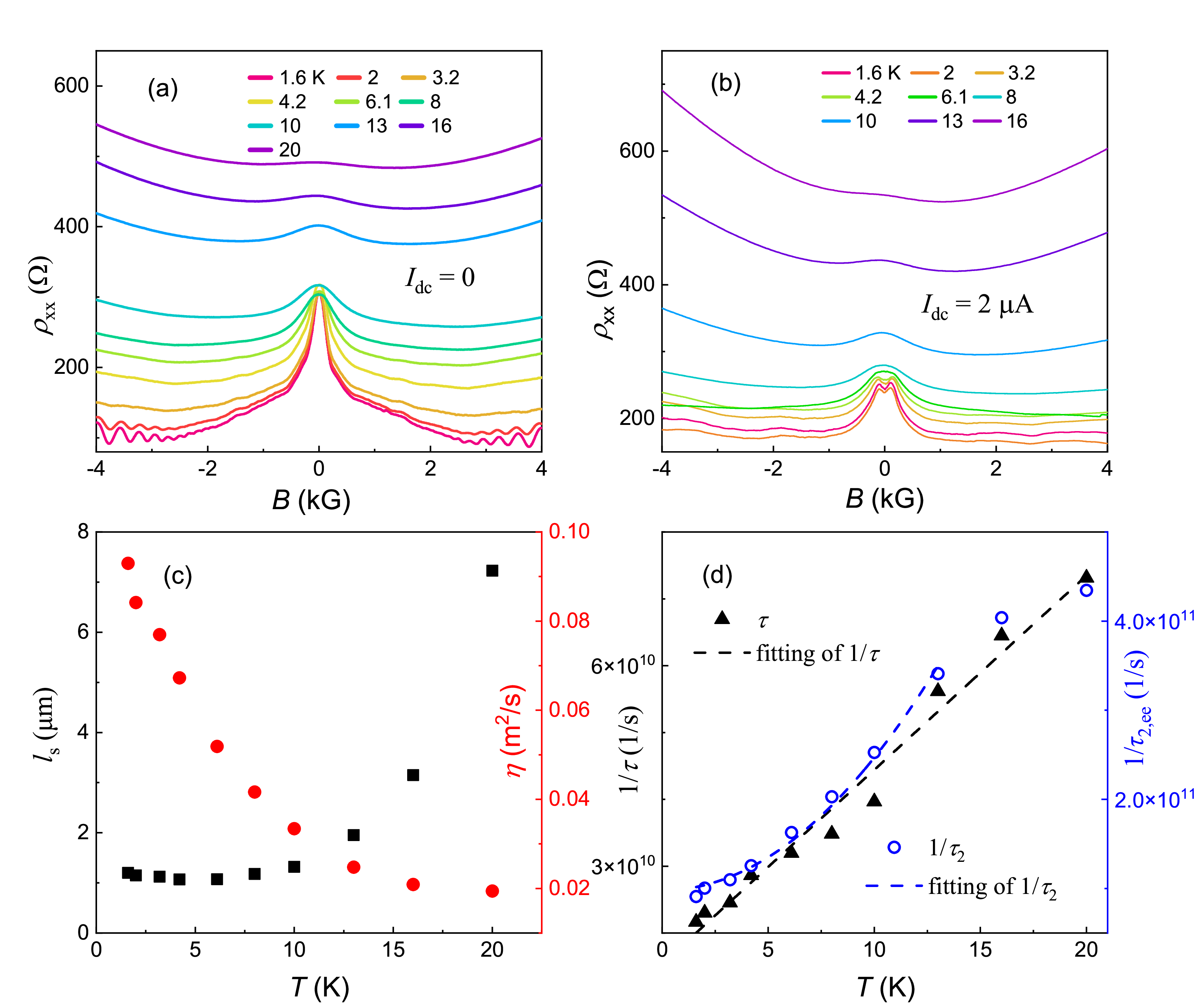}
\caption{\label{Fig4} (Color online).
Panel (a): $T$-dependent resistance with a U-turn current configuration of measurement of Structure I at zero dc-current.
(b): $T$-dependent resistance with an external U-turn current configuration of Structure I at a bias of $I_{dc} = 2\ \mu$A.
(c): $T$-dependent slip length $l_{s}$ (black solid squares) and viscosity $\eta$ (red solid circles).
(d): Calculated momentum relaxation rate $1/\tau$ (black solid triangles) and second momentum relaxation rate $1/\tau_{2}$ (red open circles) versus $T$, along with theoretical values (highlighted by dashed lines).
}
\end{figure*}

In order to further explore the physical effect of dc bias on the e-e interaction and viscosity, we investigate the $T$-dependent transport (of structure I) without dc current and with $I_{dc} = 2\ \mu$A (as shown in Fig. 4(a) and 4(b)), respectively.
In panel (b), the applied dc current induces an additional valley at $T < 6.1$ K and smears out at high $T$.
The viscosity decreases and the slip length increases monotonously with $T$ (in Fig. 4(c)).
According to theoretical and experimental studies about the Gurzhi effect \cite{gurzhi_electron-electron_1995, qian_lifetime_2005, gusev_stokes_2020}, we analyze the scattering rates $1/\tau_{2, ee}$ and $1/\tau$ by means of \begin{equation}
    \begin{aligned}
        \frac{1}{\tau_{2, ee}\left(T_e\right)} &= \frac{1}{\tau_{2, 0}} + A_{ee}^{FL} \frac{T_{e}^{2}}{[\ln (E_F/T_e)]^2} \\
        \frac{1}{\tau \left(T_{l}\right)} &= \frac{1}{\tau_0} + A_{ph}T_{l}
    \end{aligned}
\end{equation}
The extracted parameters from our $T$-dependent experiments are as follows: $1/\tau_{2,0} = 9.77 \times {10}^{11}\ s^{-1}$, $A_{ee}^{FL} = 4.4 \times {10}^{9}$ T$^{-2}$ s$^{-1}$, $1/\tau_{0} = 1.536 \times {10}^{10}$ s$^{-1}$, $A_{ph} = 2.9 \times 10^{9}$ T$^{-1}$ s$^{-1}$.
In contrast, the relaxation rates $1/\tau$ and $\tau_{2, ee}$ exhibit different features: $1/\tau$ is approximately linear to $T$, and $1/\tau_{2, ee}$ decreases slowly at around $T < 4$ K.

\section{\label{sec:3}DISCUSSION}
Our hydrodynamic transport study reveals the relation between the dc current density $j_{dc}$ and the second momentum relaxation time $\tau_{2, ee}$.
The hydrodynamic regime requires $l_{ee}< w \ll l$, where the mean free paths are approximately $l \sim 20$ $\mu$m or larger, the correlation length is $l_{ee} \sim 2.6$ $\mu$m, and the defined current width in Hallbar device is $w = 4$ $\mu$m.
At a dc current bias, the viscous resistivity drops rapidly and the viscosity dramatically weakens, which illustrates a distinct transition from hydrodynamic regime to Ohmic regime \cite{PhysRevB.104.195415}.
The phase transition in electron liquids only driven by drift velocity or dc-current are stable at vairous circumstances, such as, at a reasonable regime of $T$ or $B$.

In order to further explain the relation between $j_{dc}$ and the electron relaxation times, we investigate $T$-dependent measurements at the dc current bias.
The second momentum relaxation rate $1/\tau_{2,ee}$ shows a quasi-quadratic dependence on $j_{dc}$ (in Fig. 3(a)), which is very similar to its dependence on $T$ (as shown in Fig. 4(d)).
Driven by an additional dc current, which is caused by an $E$ field ($E_x$), a rigorous derivation of the modified electron distribution occurs.
Under the assumption that the electron distribution function is spatially uniform, the Boltzmann's equation can be expressed as:
\begin{eqnarray}
\frac{\partial f}{\partial t}+\frac{-e}{m}\boldsymbol{E}\cdot\nabla_{k} f = -\frac{f - f_{0}}{\tau_{e}} = C_{ee}(f)
\end{eqnarray}
, where $C_{ee}(f)$  as a function of $f$ represents the collision terms of electron-electron scattering.
In order to solve the steady-state solution with $\frac{\partial f}{\partial t}=0$, the equation becomes $\frac{-e}{m} \boldsymbol{E} \cdot \nabla_{k} f = - \frac{f - f_{0}}{\tau_{e}} = C_{ee}(f)$.
At a small $E$-field with $\frac{\partial f}{\partial v_{x}}\approx\frac{\partial f_{0}}{\partial v_{x}}$, the approximated solution can be derived as a linear term of $E$.
However, in our experiments, the assumption of a small $E$-field may not hold, as the electron fluid model generates a non-uniform $E$-field and induces a non-uniform drift velocity distribution of electronic liquids.
Since the collision term ($C_{ee} (f)$) involves a sophisticated nonlinear integral form, it is difficult to deduct analytical solutions and obtain a distinct physical picture.
To address this complexity, we adopt a simplified model based on ``overheated electrons".
At a sufficiently low $T$ of several Kelvins, the Fermi-Dirac distribution of electrons approximates a step function.
At a relatively large $E$-field, the distribution of electron states evolves from Fermi-Dirac distribution into a new equilibrium state characterized by the broadening of energy redistribution and can be described by $T_e$, which is usually unequal to the $T_{l}$ of samples.
Consequently, the electron liquid (state) obeys the thermal equilibrium condition of $P_{in} = P_{out}$ with the expressions:
\begin{equation}
\begin{aligned}
    P_{in}&=\rho j^{2}\\
    P_{out}&=\frac{C_{e}\left(T_{e}-T_{l}\right)}{\tau_{e-ph}}=\frac{{\gamma T_{e}}\left(T_{e}-T_{l}\right)}{\tau_{e-ph}}
\end{aligned}
\end{equation}

\begin{figure}
    \centering
    \includegraphics[width=\linewidth]{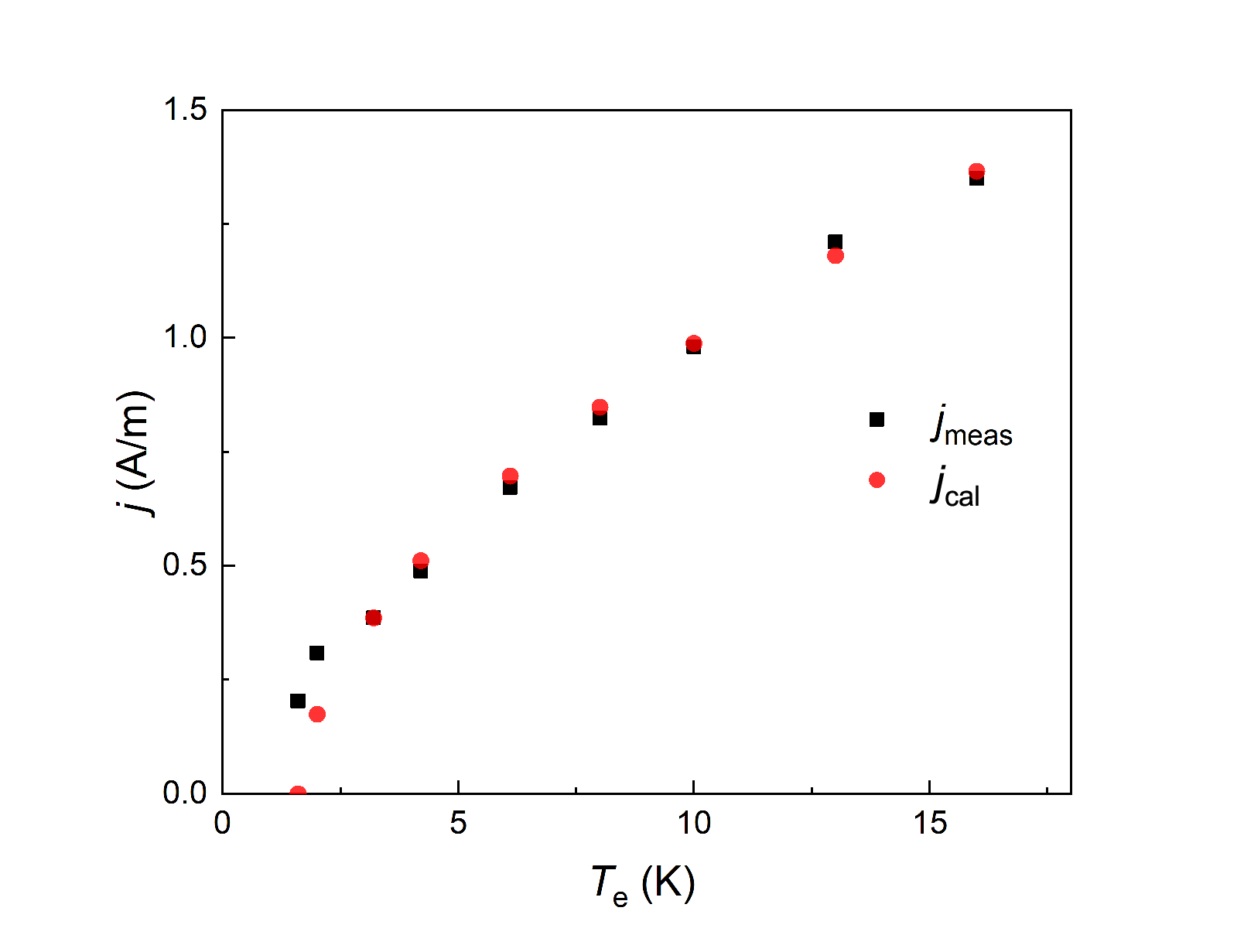}
    \caption{(Color online).
    The equivalent current densities at various temperatures (black solid squares) are estimated via equal second momentum rates with Eq. (5) and (8), and the heating effect from $j_{dc}$ versus $T_{e}$ (red solid circles) is deduced with Eq. (11).}
    \label{Fig5}
\end{figure}
In the equations, $P_{in}$ represents the input heat flow caused by $E$-field bias (Joule heat), and $P_{out}$ is the heat flow transfer between electrons and phonons characterized by electron-phonon energy relaxation time $\tau_{e-ph}$ \cite{ma15093224}.
$C_e$ is the heat capacity of electrons, and the coefficient $\gamma$ is defined as the ratio between the heat capacity and the $T$ of electron channel, based on Sommerfeld Model \cite{GRIVEI199890}.
In a highly degenerate 2DEG with a Fermi energy of $T_{F} \sim 82$ K, $\gamma$ is calculated via
$\gamma = \frac{\pi^{2} k_B^{2}}{3}D_{2D} \left(E_{F}\right) = \frac{\pi k_{B}^{2} m^{\ast}}{3 \hbar^{2}} \approx 1.09 \times {10}^{-9}$ J$\cdot$ m$^{-2} \cdot$ K$^{-2}$.
The e-phonon relaxation time $\tau_{e-ph}$ is estimated from ref. \cite{ma15093224}, and the expression of $j$ is deduced as:
\begin{eqnarray}
j = \sqrt{\frac{{\gamma T_{e}} \left( T_{e} - T_{l} \right) }{{\rho(T_{e}) \tau_{e-ph}}}}
\label{eq11}
\end{eqnarray}
, where the probed resistivity $\rho$ is a function of $T_{e}$.
(Because $\rho \approx \rho_{0} (1/\tau + 1/\tau^{\ast})$ and $\tau^{\ast}$ is a function of $T_{e}$, and the simplified theory and the detailed parameters are displayed in Section II and Table I of the Supplementary Information).
In order to clarify the effect from the dc current bias, we calculate the corresponding current density $j$ (as shown in red solid circles in Fig. 5) via Eq. (11) with a reasonable $\tau_{e-ph} = 11.3$ ps.
The experimental values of dc-current $j_{meas}$ (black solid squares in Fig. 5) are obtained from fitted data of $j_{dc}$ and $T_{e}$.
Since $\tau_{2, ee}$ is a function of electron temperature (Eq. (8)) and is tunable via dc-current (Eq. (5)), we obtain the corresponding $j_{cal}$ vs. $T_{e}$ (in Fig. 5) by means of bridging the two equations.
The experimental $j_{dc}$ is consistent with the theoretical values at $T > 3$ K, although discrepancy occurs at very low $T$ (e.g. $T \sim 1.5$ K).
So we conclude that the heated electron scattering dominates in the magnetoresistance experiments with dc-current, and due to the Gurzhi effect, the negative magnetoresistance is quite sensitive.
The sensitivity stems from the characteristic quadratic $T_{e}$-dependent e-e scattering rate, i.e. $1/\tau_{2, ee} \sim {T_e}^2$.
Thus at $B = 0$, the viscous resistivity $\Delta \rho \sim \eta \sim 1/\tau_{2, ee} \sim T^{-2}$ is consistent with the Gurzhi effect in electron liquid \cite{gurzhi_electron-electron_1995}.
Therefore, monitoring the transport under a dc current provides an effective methodology for probing $T_{e}$ in ultra-clean and e-e interaction-dominated electron systems.

\begin{figure}[ht]
    \centering
    \includegraphics[width=\linewidth]{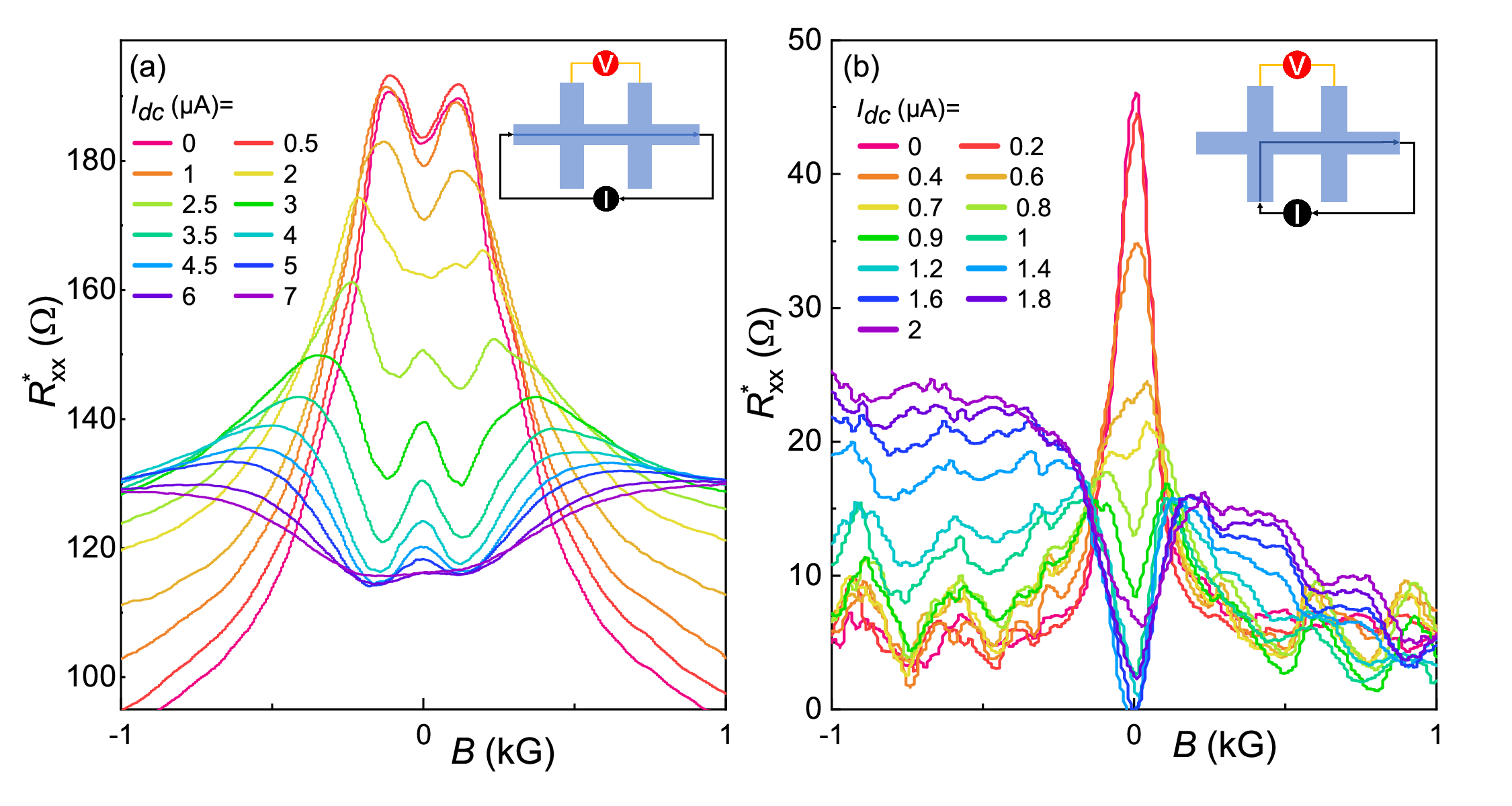}
    \caption{(Color online).
    Panel (a): DC-bias dependent transport (with ordinary configuration of measurement) of structure II at 1.6 K.
    (b): DC-bias dependent transport (with L-shape current configuration measurement) of structure IV at 1.6 K.}
    \label{Fig6}
\end{figure}

Due to the boundary condition dependence of hydrodynamic transport, the measured features exhibit differences between configurations of measurement.
In Fig. 6(a), distinct from the typical Lorentzian peak, the ordinary configuration measurement (from source to drain) in a Hallbar displays a Lorentzian peak superposing with an additional minimum at $B = 0$ (exhibiting a ``M"-curve).
And similar to our results driven by U-turn current, at high $j_{dc}$ the Lorentzian valley occurrence is accompanied by an additional maximum (``W"-curve ) at zero $B$.
In comparison, at a dc current of a L-loop (input from a side contact to a drain contact on $x$-axis, as shown in the inset of Fig. 6(b)), a Lorentzian peak occurs at zero dc-bias and evolves into a valley at a high $j_{dc}$.
In addition, oscillating features from the various orders of cyclotron resonances (CR) are observed
\cite{ji_negative_2021}.

Despite its great potential in hydrodynamic electron fluids, the development of achievable devices still faces challenges.
The simulation of hydrodynamic electron fluids resembles water flow in liquid, so that it is very sensitive to boundary condition and electron properties such as Fermi energy, effective mass and mobility.
We expect further research on the boundary condition and accurate simulation to address this issue.

\section{\label{sec:4}CONCLUSION}
In our investigation of the differential resistances of electron liquid in a 2DEG, we observe unusual dc-current dependent transitional Lorentzian features at very low $B$-fields.
By analyzing the transport results via the hydrodynamic electron theory, we obtain the dc-current dependent scattering rates $1/\tau$ and $1/\tau_{2, ee}$, the latter of which is proportional to the viscosity $\eta$.
And the extracted $T$-dependent viscous resistivity ($\Delta \rho$) exhibits a universal Gurzhi effect in various measurements.
Moreover, a hidden relation between the electron temperature and input dc current is established from the e-e interaction dominated hydrodynamic transport traces, which is consistent with the estimated electron temperature method via quantum Hall traces.
The method we developed is effective for probing or monitoring $T_{e}$ in electron liquids with a wide $T$ range (from high to low temperatures), and it may become a more general approach to detect $T_{e}$ via transport.
Our results may also stimulate theoretical characterization and experimental research about the differential transport of electron liquids.

\section{\label{sec:5}Acknowledgments}
We would like to thank Changli Yang for the guidance work on the wafer growth.
This project is supported by the National Science Foundation of China (Grant No.11974339, and 12574213), by the Strategic Priority Research Program of the Chinese Academy of Science (Grant No. XDB 0460000), and by the Quantum Science and Technology--National Science and technology Major Project (Grant No. 2021ZD0302600).
Y.W. and C.Z. performed experiments; C.Z. and Y. W. analyzed data and wrote the paper; Y.W. carried out the cleanroom work; S.Z. and L.L. grew the high-quality semiconductor wafers; C.Z. conceived and supervised the project.

\bibliography{hydrodynamic}

\end{document}